\documentclass[pra,showpacs,floatfix,amsmath,amsfonts,
twocolumn
]{revtex4}

\usepackage{amssymb,amsmath}
\usepackage{graphicx}
\usepackage{eurosym}
\usepackage{color}
\usepackage[normalem]{ulem}
\usepackage[makeroom]{cancel}

\newcommand{\p}{{\cal P}}
\newcommand{\T}{{\cal T}}
\newcommand{\PT}{\mathcal{PT}}
\newcommand{\C}{\mathcal{C}}
\newcommand{\CPT}{\mathcal{CPT}}

\newcommand{\bp}{{\bf p}}
\newcommand{\bq}{{\bf q}}

\newcommand{\hp}{{\hat{p}}}

\newcommand{\hK}{{\hat{K}}}

\begin{document}

\title{$\CPT$-symmetric coupler with  intermodal dispersion}

\author{Dmitry A. Zezyulin$^{1}$, Yaroslav V. Kartashov$^{2,3,4}$, and Vladimir V. Konotop$^{1}$}

\affiliation{$^{1}$Centro de Fisica Te\'orica e Computacional and Departamento de F\'isica, Faculdade de Ci\^encias,
	Universidade de Lisboa,  Campo Grande 2, Edif\'icio C8, Lisboa 1749-016, Portugal\\
	$^{2}$ICFO-Institut de Ciencies Fotoniques, The Barcelona Institute
	of Science and Technology, 08860 Castelldefels (Barcelona), Spain\\
	$^{3}$Institute of Spectroscopy, Russian Academy of Sciences, Troitsk,
	Moscow Region, 142190, Russia\\
	$^{4}$Department of Physics, University of Bath, BA2 7AY, Bath, United Kingdom}

\date{\today}

\begin{abstract} 
	
	A dual-core waveguide with balanced gain and loss  in different arms 
	and with intermodal coupling is considered. The system is not invariant under the conventional  $\PT$ symmetry but obeys $\CPT$ symmetry where an additional   spatial inversion $\C$ 
	corresponds to  swapping  the coupler arms. We show  that second-order dispersion of coupling allows for unbroken $\CPT$ symmetry and supports  
	propagation of stable vector solitons along the coupler.  Small-amplitude solitons are found in  explicit form. The combined effect of  gain-and-loss and  dispersive coupling results in several interesting features which include a separation  between the components in different arms, nontrivial dependence of stability of a soliton on its velocity, and the existence of more complex stationary two-hump  solutions. Unusual decay dynamics of unstable solitons is discussed too.
	
\end{abstract}



\maketitle

Physical properties of the mode coupling in parallel optical  waveguides determine functionalities of  devices based on such structures. One of the mentioned properties is the dispersion of the coupling, which was experimentally observed in~\cite{ChiangExper}. When waveguides are nonlinear, intermodal coupling affects the existence and stability of two-component optical solitons. This topic has been initiated by studies in  \cite{Chaing97a,Chaing97b} and received considerable attention in the theory of nonlinear couplers. In particular, there have been performed detailed numerical studies of soliton propagation and switching in conservative directional couplers whose arms obey Kerr nonlinearity with account of first- \cite{RaChiAk} and second-order \cite{RaPa} dispersive terms  in coupling constant. Selective all-optical switching was described in~\cite{Wang}. Peculiar dark solitons supported by dispersive coupling were reported in~\cite{KaMaKo}. More recently it was found that dispersive coupling stabilizes spatio-temporal solitons in Kerr media~\cite{KMKLT}. 

In this Letter, we consider a dual-core waveguide with balanced gain and loss and show that this system supports propagation of stable linear waves and solitons whose properties are significantly affected by the   intermodal dispersion. The dual-core waveguide  is governed by the system 
\begin{eqnarray}
\label{eq:nls}
\begin{array}{l}
\displaystyle 	i \frac{\partial q_1}{\partial z} = - \frac{\partial^2 q_1}{\partial \tau^2}  +i\gamma q_1 - Kq_2  -2|q_1|^2q_1,
\\[1mm]
\displaystyle 	i \frac{\partial q_2}{\partial z}  = -  \frac{\partial^2 q_2}{\partial \tau^2}  - i\gamma q_2 - Kq_1  -2|q_2|^2q_2.
\end{array}
\end{eqnarray}
Here $q_{1,2}$  
are  the dimensionless fields,  
$z$ and $\tau$ are the propagation coordinate and reduced time, respectively, and $\gamma>0$ 
describes gain in the first and dissipation in the second waveguides. The intermodal dispersion is described by the operator~\cite{Chaing97a,Chaing97b,RaPa}
$
K = \kappa_0 + i\kappa_1\partial_\tau - \kappa_2\partial^2_\tau,
$
with real constants $\kappa_{0,1,2}$ characterizing dispersionless coupling ($\kappa_0$), first ($\kappa_1$) and second ($\kappa_2$) orders of the dispersion.  These coefficients may vary in relatively wide range as functions of the carrier frequency, width of the  pulse, and material parameters \cite{Chiang95,RaPa}.

At $\gamma=\kappa_{1,2}=0$ and $\kappa_0\neq0$ the model is reduced to the  
conservative unidirectional coupler~\cite{Wright}.  Nonzero $\gamma$ corresponds to a system with   balanced gain and losses. For $\kappa_{1,2}=0$,  
double-core structures of this type, have been attracting steady 
 attention in the context of $\PT$ symmetry \cite{BenderBoet},  without~\cite{YK2,YK3} and in the presence~\cite{DribMal,AKOS} of temporal dispersion (see \cite{review} for a comprehensive review).      
When $\kappa_2=0$ and  other parameters are nonzero, \eqref{eq:nls} is reduced to the model of $\mathcal{CPT}$-symmetric spin-orbit coupled Bose-Einstein condensate, introduced in~\cite{KKZ} and considered in the two-dimensional setting in~\cite{MalSak}.

Following~\cite{KKZ}, we  notice that system (\ref{eq:nls}) is not invariant under the conventional $\PT$ transformation with the ``time''-reversal operator $\T$: $i\to-i$, $z\to -z$ and the ``spatial" inversion operator $\p$ defined as in Quantum Mechanics $\tau\to-\tau$~\cite{BenderBoet} (notice that in optical applications the roles of time and space coordinates are exchanged). 
Also, the model (\ref{eq:nls}) is not invariant under the action   $\C\T$ transformation, where operator $\C$ is a transverse spatial inversion, defined as in optics of guided structures~\cite{YK2,YK3}, by swapping the waveguides $q_1\leftrightarrow q_2$. However model (\ref{eq:nls}) is symmetric if all three symmetry operators are involved, i.e., it is $\CPT$ symmetric. We notice that the usage of  notation $\C$ stems from the similarity of this transformation with the charge operator (see Refs.~\cite{KKZ,review} for more details). For discussion of the operator  $\C$ in   another double-core system with gain and losses we mention Ref.~\cite{CP_optics}.

\begin{figure}
 	\includegraphics[width=\columnwidth]{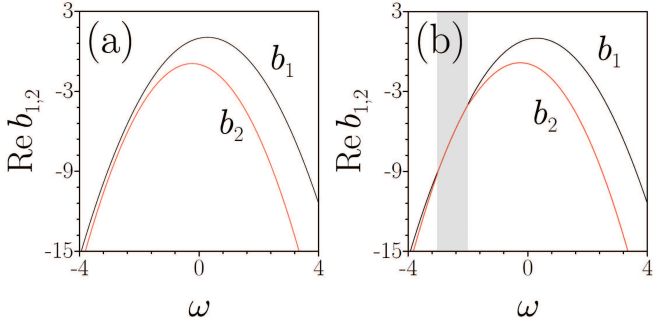}
 	\caption{Linear dispersion relations for   $\kappa_{\{0,1,2\}} = \{1, 0.5, 0.1\}$ and (a) $\gamma=0.2$ (unbroken $\CPT$-symmetric phase), and (b)  $\gamma=0.4$ (slightly above the  $\CPT$ symmetry breaking threshold $\gamma_{\CPT}=0.375$). Shaded domain in (b) shows the frequency band where the dispersion curves merge and the corresponding propagation constants become complex-valued.}
 	\label{fig:one}
 \end{figure}

Without loss of generality, we assume that $\kappa_0,\kappa_1>0$. The linear dispersion relation, $q_{1,2}\propto e^{ibz - i\omega\tau}$, of 
Eq.~(\ref{eq:nls}) consists of  two branches:  
$
{b}_{1,2} = -\omega^2  \pm (\hK^2(\omega)-\gamma^2)^{1/2},
$ 
where  
$
\hK(\omega) = \kappa_2\omega^2 + \kappa_1\omega+\kappa_0.
$
Thus the linear spectrum is all-real if $\kappa_1^2 < 4\kappa_0\kappa_2$ and the gain-and-loss coefficient is below the $\CPT$-symmetry breaking threshold:
$
\gamma \leq \gamma_{\CPT}= \kappa_0 - {\kappa_1^2}/{(4\kappa_2)}.
$
Thus the spectrum always contains  complex propagation constants for $\kappa_2=0$. However, sufficiently strong second-order dispersion $\kappa_2>0$ allows to achieve unbroken   $\CPT$ symmetry, i.e., makes the spectrum all-real. Examples of dispersion curves for the system with the unbroken and broken  $\CPT$ symmetries are presented in Fig.~\ref{fig:one}. If the former case [Fig.~\ref{fig:one}(a)] both propagation constants are real for any $\omega$. For the broken $\CPT$ symmetry [Fig.~\ref{fig:one}(b)], there is a band of frequencies [shown by a gray stripe] where the two dispersion curves merge, and eigenvalues   acquire nonzero imaginary parts.

System (\ref{eq:nls}) 
can be significantly simplified in the absence of resonant processes, as explained below. Indeed, let us perform the Fourier transform of (\ref{eq:nls}) with respect to $\tau$,    introduce the rotation matrix 
\begin{eqnarray}
S(\omega)=
\left(\!\! \begin{array}{cc}
e^{-i\alpha(\omega)} & -e^{i\alpha(\omega)}
\\
e^{i\alpha(\omega)} & e^{-i\alpha(\omega)}
\end{array}\!\!\right), \quad \alpha(\omega) = \frac{1}{2}\arcsin\left[\frac{\gamma}{\hK(\omega)}\right], 
\end{eqnarray}
and define the column-vectors $\bq=(q_1,q_2)^T$ (``$T$'' stands for transpose), and $\hat{\bp}=S^{-1} \hat{\bq}$. Hereafter a hat-symbol stands for the Fourier transform: i.e. $\hat{q}_n=F[q_n]=\int_{-\infty}^{\infty} q_n e^{i\omega \tau} d\tau$. Then, it is straightforward to show   that $\hp_{1,2}$ solve
\begin{equation}
\label{eqp}
\begin{array}{l}
\displaystyle  \frac{\partial \hp_1}{\partial z}= {i}b_1\hp_1+\frac{i}{\cos(2\alpha)}(e^{-i\alpha}F[|q_1|^2q_1] +e^{i\alpha}F[|q_2|^2q_2]),
\\[2mm] 
\displaystyle  \frac{\partial \hp_2}{\partial z}={i}b_2\hp_2-\frac{i}{\cos(2\alpha)}(e^{i\alpha}F[|q_1|^2q_1] -e^{-i\alpha}F[|q_2|^2q_2]).
\end{array}
\end{equation}
If  solution $\hat{\bp}$ is found, the field in the coupler is given by the inverse Fourier transform $\bq=\frac{1}{2\pi}\int_{-\infty}^{\infty} e^{-i \omega \tau}S(\omega)\hat{\bp}(\omega)d\omega$. 

Now we assume that there exist no frequencies $\omega$, $\omega_{1,2}$ for which at least one of the   two resonant conditions
\begin{equation}
 b_{n}(\omega_1)+b_{n}(\omega_2)-b_n(\omega_1+\omega_2-\omega)=b_{3-n}(\omega), \quad n=1,2,
\end{equation}
is satisfied. Then one can apply the rotating wave approximation (RWA)~\cite{RWA} to equations (\ref{eqp}), i.e. neglect all terms ``rotating'' with the frequency different from $b_n$  in the Fourier integrals in the equation for $\hp_n$  ($n=1,2$). In this way one ensures that the system (\ref{eqp}) admits nontrivial solutions where either $\hp_1(\omega)\equiv 0$ or $\hp_2(\omega)\equiv 0$.  For such solutions the system of  two nonlinearly coupled equations \eqref{eqp} is reduced to a single one. As an example, we consider the case $\hp_2(\omega)\equiv 0$. Then \eqref{eqp}  gives
\begin{eqnarray}
\label{eqp1_int}
i\frac{\partial \hp_1(\omega)}{\partial z}= {-}b_1(\omega)\hp_1(\omega)- \iint d\omega_1d\omega_2 G(\omega,\omega_1,\omega_2)
\nonumber \\ \times
\hp_1(\omega_1) \hp_1(\omega_2)\hp_1^*(\omega_1+\omega_2-\omega),
\end{eqnarray}
where 
\begin{equation*}
G(\omega,\omega_1,\omega_2)= \frac{\cos\left[\alpha (\omega_1)+\alpha (\omega_2)-\alpha(\omega_1+\omega_2-\omega)  +\alpha(\omega)\right]}{2\pi^2\cos[2\alpha(\omega)]}.
\end{equation*} 

Several comments are in order here. First,  the derivation of \eqref{eqp1_int}  does not rely on the specific form of $\hK(\omega)$. Therefore this equation takes into  account  dispersion of all orders accounted by the coupling $K$ and remains valid for a general integral coupling of the form   $Kq=\int {\cal K}(\tau-t)q(t) dt$ where ${\cal K}(\tau)$ is some localized function. 

Second, the solution $\hp_1\not\equiv 0$, $\hp_2\equiv 0$ (or {\it vice versa}) describes solutions  with conserved   total energy flow  $U=\int
\left(|q_1|^2+|q_2|^2\right)d\tau$: $dU/dz=0$. This follows from the Parceval equality  $\int
|q_1|^2d\tau  = \int
|\hp_1|^2d\omega  =	\int
|q_2|^2d\tau$.  

Third, in spite of the approximate character of Eq.~(\ref{eqp1_int}) obtained under the RWA, 
it gives an \textit{exact} plane wave solution of the original system (\ref{eq:nls}). Indeed, a particular solution of (\ref{eqp1_int}) in the form $\hp_1 = \sqrt{2}\pi A \exp\{i  [b_1(\Omega) +|A|^2]z\}\delta(\omega-\Omega)$, where $A$ and $\Omega$ are arbitrary amplitude and frequency, respectively, corresponds to the plane  wave   
\begin{equation}
\label{eq:plain}
q_{1,2} = ({A}/{\sqrt{2}})\exp\{ i [b_1(\Omega)+|A|^2]z-i\Omega\tau  \mp  i \alpha(\Omega)\}.
\end{equation}

\begin{figure}
	\centering
		\includegraphics[width=\columnwidth]{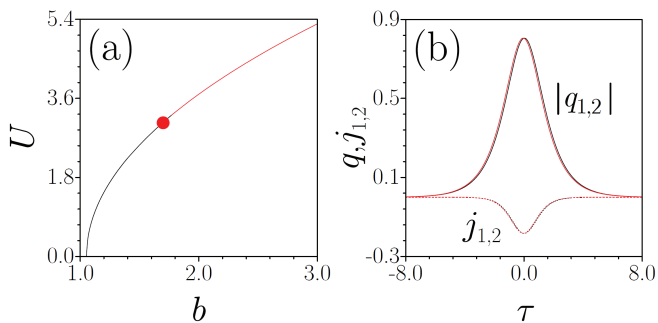}%
	\caption{(a) Family of 
		solitons for $\gamma=0.2$ and $\kappa_{\{0,1,2\}}=\{1, 0.5, 0.1\}$. Solutions are stable for $b\lesssim 1.78$ and unstable otherwise. (b) Example of a stable soliton from (a) corresponding to  $b=1.7$.  Solid and dashed  curves show moduli $|q_{1,2}|$ current distributions  $j_{1,2}$, respectively.}
	\label{fig:two}
\end{figure}

Finally, \eqref{eqp1_int} is a convenient starting point for obtaining approximate small-amplitude solitons.  To this end, we consider the  frequency $\omega_0$  corresponding to the maximum of the upper  branch of the linear spectrum, i.e., the frequency defined from $b_1^\prime (\omega_0)=0$ (hereafter $b_j'(\omega)\equiv{\partial b_j}/{\partial \omega}$). We assume that $\hp_1(\omega)$ is a small-amplitude function  
well-localized around $\omega=\omega_0$. Performing  the Taylor expansion of Eq.~(\ref{eqp1_int}) in powers of $\omega-\omega_0$ and calculating the inverse Fourier transform, we arrive at the standard nonlinear Schr\"odinger (NLS) equation for the field $p_1$: 
\begin{equation}
\label{NLS}
\frac{\partial p_1}{\partial z} = i\left(b_1+\frac{\omega_0^2 b_1''}{2}\right)p_1 + \omega_0b_1''\frac{\partial p_1}{\partial\tau}- \frac{ib_1''}{2} \frac{\partial^2 p_1}{\partial\tau^2} + 2i|p_1|^2p_1.
\end{equation}
Here $b_1$ and $b_1''$ are computed at $\omega=\omega_0$. Equation (\ref{NLS}) has a solution in the form of the bright soliton. In terms of the original field variables it reads
\begin{eqnarray}
\label{eq:sym}
q_{1,2} = a\frac{\exp\left\{ i[a^2+b_1(\omega_0)] z-i \omega_0\tau\mp i\alpha(\omega_0)\right\}}{\cosh \left[a ({2/|{b}''_{1}(\omega_0)|})^{1/2} \left(\tau \pm \alpha'(\omega_0)\right)\right]}, 
\end{eqnarray}
where  $a$ is arbitrary (small) amplitude. Thus, the combined effect of the gain and loss and the dispersive coupling introduces temporal  shift $2\alpha'(\omega_0)$ between the soliton components. In order to estimate this shift,  it is  convenient to notice  that  for sufficiently small first-order dispersion $\kappa_1$ 
the frequency $\omega_0$ for which the dispersion curve $b_1(\omega)$ attains its maximum can be approximated as $\omega_0 \approx {\kappa_1 \kappa_0}/[2((\kappa_0^2-\gamma^2)^{1/2} - \kappa_0\kappa_2)]$. For example, 
for the parameters in   Fig.~\ref{fig:one}(a) and in Fig.~\ref{fig:two}, $\omega_0\approx 0.28$.  Using this value together with the set of representative parameters   Fig.~\ref{fig:two}, we compute $\alpha'(\omega_0)\approx -0.043$, i.e., the field in the arm with gain (loss) is slightly shifted to the domain $\tau>0$  ($\tau<0$).  
Another effect of  the dispersive coupling  consists in    non-zero energy currents  $j_{n}=|q_{n}|^2(\arg q_{n})_\tau$ in   the soliton components [illustrated in Fig.~\ref{fig:two} (b)].  Eq.~(\ref{eq:sym}) gives a simple estimate   $j_{n}\approx  -\omega_0|q_{n}|^2$.

In the limit of nondispersive coupling, $\kappa_1=\kappa_2=0$, solution (\ref{eq:sym}) reduces to symmetric \cite{DribMal} (or ``high-frequency'' \cite{Barashenkov}) soliton,   and upon further simplification of the model corresponding to $\gamma=0$ it reduces to the symmetric   mode $q_1=q_2$ in the coupled NLS  equations~\cite{Wright}.

For solitons of arbitrary amplitude, we use the substitution $q_{n}=e^{ibz}w_{n}(\tau)$. In the vicinity of the  small-amplitude limit, the propagation constant $b$ and functions $w_{n}$ can be approximated by the   expression (\ref{eq:sym}). Then the soliton family can be numerically continued up to    arbitrary  amplitude. An example of the soliton family and a representative  profile are shown in Fig.~\ref{fig:two}.  Using the substitution $q_j=e^{ibz}[w_n + f_n(\tau)e^{\lambda z} + g_n^*(\tau) e^{\lambda^* z}]$, where $f_n,g_n$ are the small perturbations and $\lambda$ is the perturbation growth rate, we have performed the linear stability analysis for the obtained solutions. 
The family shown in Fig.~\ref{fig:two} is stable for sufficiently  small amplitudes; 
at $b\approx 1.78$ the solutions become unstable due to appearance of a pair of purely real  eigenvalues $\lambda =\pm \lambda_0$ indicating     strong exponential instability.

Proceeding in a similar way, one can construct a family of solitons  bifurcating at the maximum of the lower dispersion curve  $b_2(\omega)$. These solitons  co-exist with the spectrum of  linear modes of the upper branch $b_1(\omega)$.  In the limit $\kappa_{1,2}=0$  they reduce  
 to   ``low-frequency'' solitons in the $\PT$-symmetric coupler   which are known to be always unstable \cite{Barashenkov}. Our results show that solitons of this type remain unstable for nonzero $\kappa_{1,2}$, and we   therefore   do not consider these solutions in this Letter. 

Now we investigate   the effect of  dispersive coupling on  moving solitons. The latter  can be obtained 
in the frame moving with a constant velocity $v$ using the Galilean transformation: $q_n=e^{i(b-v^2/4)z+iv\tau/2}w_n(\tau-vz)$. The  
fields $w_{1,2}$ solve the system  ($n=1,2$)
\begin{equation}
 \frac{\partial^2 w_n}{\partial\xi^2}-[b-(-1)^n]w_n+K_v w_{3-n}+2|w_n|^2w_n=0,
\end{equation}
where  $\xi = \tau - vz$, and the coupling is modified  by the velocity:
\begin{equation}
\label{eq:Kv}
K_v = \kappa_0 - {v}\kappa_1/2 + {v^2}\kappa_2/4 + i(\kappa_1 - \kappa_2 v)\partial_\xi -  \kappa_2\partial^2_\xi.
\end{equation}
Thus the nonzero velocity effectively changes the coupling coefficients $\kappa_0$ and $\kappa_1$, thereby affecting   the soliton's properties. 
 
\begin{figure}[t]
	\includegraphics[width=\columnwidth]{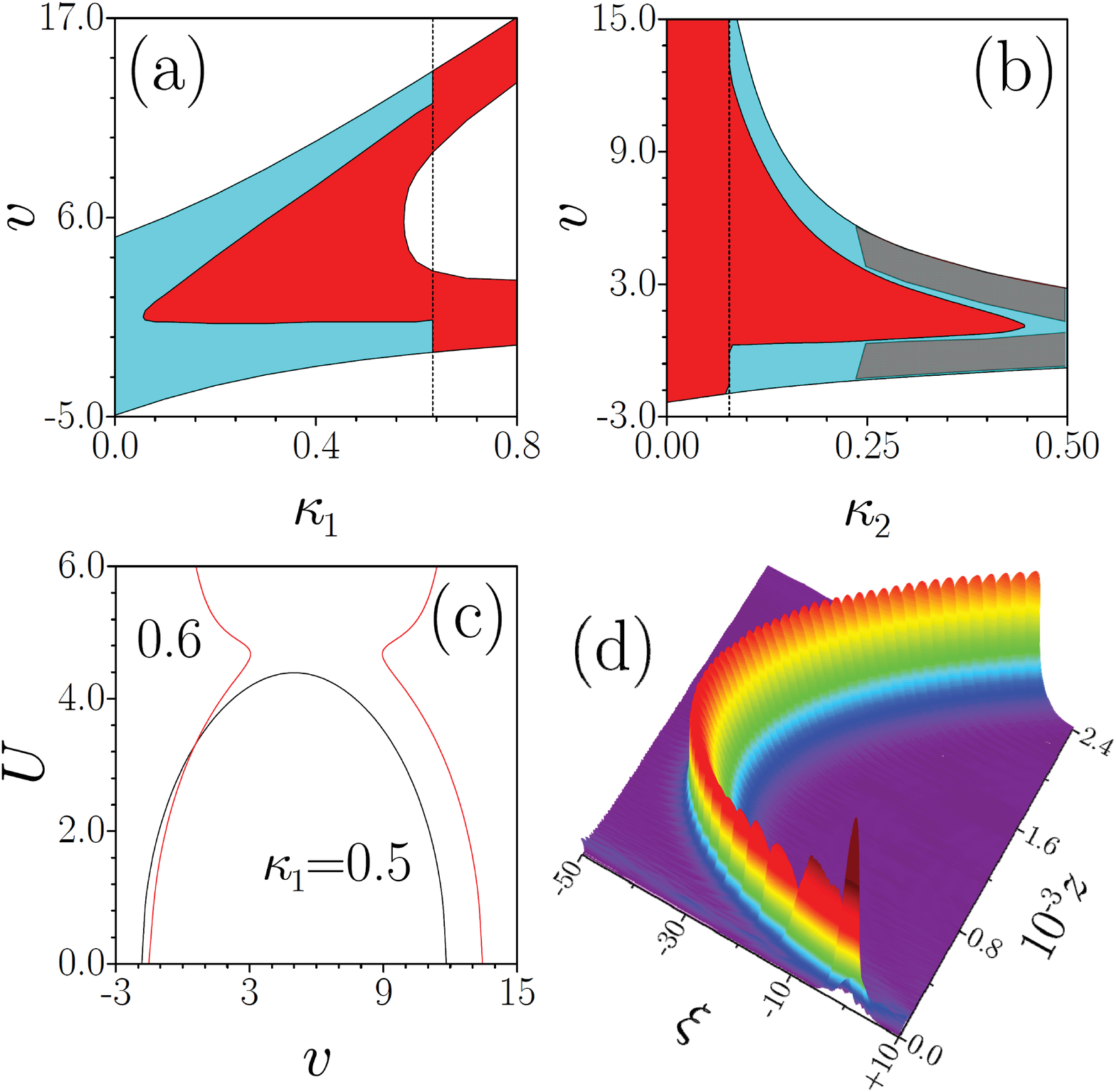}
	\caption{Domain of existence (colored) of symmetric solutions  with $b=1.66$ (a) on the plane $(\kappa_1, v)$ for $\kappa_2=0.1$, and (b) on the plane $(\kappa_2, v)$ for $\kappa_1=0.5$. Vertical  dashed lines demarcate the domain of unbroken $\CPT$ symmetry. Cyan domains correspond to stable solitons; red domains correspond to solitons with relatively strong (exponential) instability and domains with broken $\CPT$ symmetry; gray domains  correspond to solitons with relatively weak (oscillatory) instabilities.  (c) Dependencies  $U(v)$ for solutions with $\kappa_1=0.5$ and $\kappa_1=0.6$ and $b$, $\kappa_2$   as in panel (a).  
	 (d) Propagation of the unstable soliton with $b=1.08$, $\kappa_1=0.5$, $\kappa_2=0.1$ and initial velocity $v=3.1$. The result is shown in the  frame $\xi$ moving with velocity $v$.   Only the first component $|q_1|$ is shown. In all panels  $\kappa_0=1$ and $\gamma=0.2$.}
	\label{fig:three}
\end{figure}

\begin{figure}[t]
	\includegraphics[width=\columnwidth]{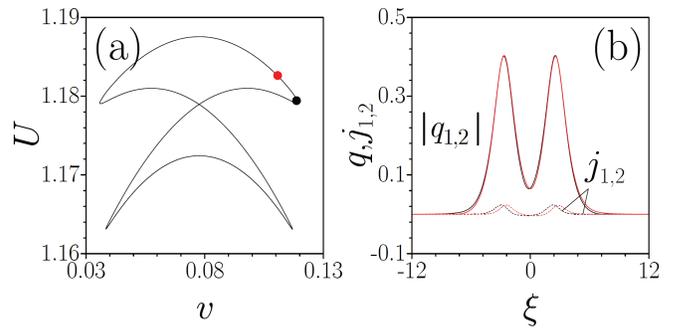}
	\caption{(Color online)  (a) Example of a  diagram showing  dependence of the power $U$  on  the soliton's velocity $v$ for two-hump solitons. Here $\kappa_0=1$, $\kappa_1=0.066$, $\kappa_2=0.85$; $\gamma=0.4$, and $b\approx 1.09$.  Red and black dots label unstable and stable solutions whose propagations are shown in Fig.~\ref{fig:five}(a) and (b), respectively. (b) Amplitudes $|q_{1,2}|$   and currents $j_{1,2}$  for the stable solution shown  in (a) with the black dot.}
	\label{fig:four}
\end{figure}

In Fig.~\ref{fig:three}(a) we show existence and stability domains of moving solitons for different strengths of the intermodal dispersion $\kappa_1$ and velocities $v$ (with other parameters kept unchanged).  Solutions exist in a finite band of velocity values.  At the boundaries of the existence region the power vanishes and the soliton transforms into linear wave [see the power curves in Fig.~\ref{fig:three}(c)]. For fixed $\kappa_1$,  the power is a non-monotonic function of $v$  approaching its maximum at $v_{\rm max}=\kappa_1/\kappa_2$. This is consistent with the change, at $v_{\rm max}$, of the sign of the first-order dispersion in the effective coupling (\ref{eq:Kv}).  As follows from the stability diagram, solutions with small power are stable, whereas   large-amplitude solitons feature strong exponential instabilities. For large $\kappa_1$ ($\gtrsim 0.55$) the  soliton family merges with a new  
family which consist of high-power solitons   [compare  curves with $\kappa_1=0.5$ and $\kappa_1=0.6$ in Fig.~\ref{fig:three}(c)]. As a result, the existence domain in Fig.~\ref{fig:three}(a) has a ``gap''.  Solutions from the new family have   two-hump profiles  [see Fig.~\ref{fig:four}(b) and the discussion below].
We also notice that the solitons   exist in the domain of broken $\CPT$ symmetry [$\kappa_1 \gtrsim 0.63$ in Fig.~\ref{fig:three}(a)]. However, they are unstable due to the instability of the linear waves.

Figure~\ref{fig:three}{(b)} illustrates the existence and stability domains for  solitons  on the plane $(\kappa_2, v)$. In this diagram we observe that apart from the domains of exponential instability (associated with  a pair of purely real eigenvalues $\pm \lambda_0$), some of the solutions 
feature relatively weak oscillatory instabilities [due to a quartet of complex eigenvalues $(\pm\lambda, \pm\lambda^*$)]. The existence and stability domains in $v$ shrink as $\kappa_2$ increases. On the other hand, for small $\kappa_2$ ($\lesssim 0.078$)  the $\CPT$-symmetric phase is broken, and solitons become   unstable.

Unstable  moving solitons develop various dynamical scenarios, depending on the system parameters. 
A relatively weak instability manifests itself in a slight change of the soliton's velocity.
A stronger  instability can lead to a more complicated propagation
when velocity of the wavepacket strongly changes with distance $z$, 
see Fig.~\ref{fig:three}(d). Notice that the dynamics    in Fig.~\ref{fig:three}(d) is  plotted  in the  frame $\xi$ moving with the constant velocity $v$ equal to the initial velocity of the soliton. Respectively, a wave moving towards $\xi\to -\infty$ in Fig.~\ref{fig:three}(d) has the velocity smaller than $v$, whereas propagation towards $\xi \to +\infty$ corresponds to velocities exceeding $v$.  Another unstable regime observed in simulations (not shown here) is the unbounded growth of the soliton  power in the active waveguide.

\begin{figure}[t]
	\includegraphics[width=\columnwidth]{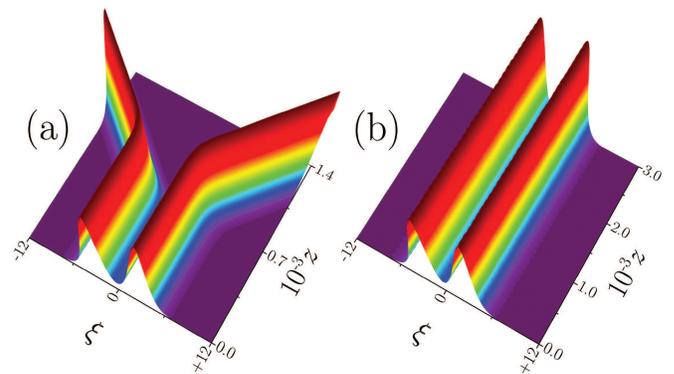}
	\caption{Dynamics  of two-hump solitons shown with red (a) and black  (b)  dots in Fig.~\ref{fig:four}. The initial velocity is $v= 0.11$ (a) and $v=0.12$ (b). In both panels, the propagation is plotted in a frame moving with velocity $v$, and only the first component $|q_1|$ is shown.}
	\label{fig:five}
\end{figure}

An interesting feature of the considered model is that, apart from the simplest one-hump solitons, it admits more complex stationary solutions in the form of multi-hump solitons.
A diversity of such solutions can be very rich (depending on the characteristics of the coupling). They can form families characterized by sophisticated power curves $U(v)$, as shown in the example  in Fig.~\ref{fig:four}(a) where four two-hump solutions with different powers coexist for a given velocity $v$ (notice that the intersection point at $v\approx 0.08$ does not correspond to a bifurcation; it corresponds to two different solution with equal velocities $v$ and powers $U$). A characteristic double-hump soliton profile is shown in Fig.~\ref{fig:four}(b). Whereas  most of the solutions in  Fig.~\ref{fig:four}(a) are unstable, a detailed stability scan reveals  small segments on this diagram where the instability growth rate vanishes, and two-hump solitons  become stable. Examples of unstable and stable propagation of two-soliton solution are presented in Fig.~\ref{fig:five}(a) and Fig.~\ref{fig:five}(b), respectively. The unstable solution eventually breaks into a pair of the solitons propagating with different velocities (one of the velocities is smaller than that of the initial soliton, whereas another soliton propagates faster than the initial two-hump state). In contrast, the  stable two-soliton configuration  propagates undistorted for indefinite propagation distance,  in spite of a  perturbation added to the initial profile.

To conclude, we investigated solitons propagating in  a $\CPT$-symmetric directional coupler whose symmetry is determined by the gain-and-loss balance  and   the intermodal coupling. Strong coupling dispersion is shown to allow for unbroken $\CPT$-symmetric phase which is broken in a weakly dispersive coupler. Small-amplitude solitons were found in the explicit form, and properties of moving one- and two-hump solutions  were described.


\begin{thebibliography}{99}

\bibitem{ChiangExper} K. S. Chiang,  Y. T. Chow, D. J. Richardson, D. Taverner, L. Dong, and L. Reekie,   
Opt. Commun. {\bf 143}, 189-192 (1997).


\bibitem{Chaing97a} K. S. Chiang,   
IEEE J. Quant. Electr. {\bf 33}, 950-954 (1997).
	
\bibitem{Chaing97b}	K. S. Chiang,  
J. Opt. Soc of Am. B, {\bf 14}, 1437-1443  (1997).

\bibitem{RaChiAk} V. Rastogi, K. S. Chiang, and N. N. Akhmediev,
Phys. Lett. A {\bf 301}, 27-34 (2002)

\bibitem{RaPa} P. M. Ramos and C. R. Paiva, 
IEEE J. Quant. Electr. {\bf 35}, 983-989 (1999).

\bibitem{Wang}  Y. Wang and W. Wang, 
Appl. Phys. Lett.  {\bf 88}, 181110 (2006).

 
\bibitem{KaMaKo} Y. V. Kartashov, B. A. Malomed, V. V. Konotop, 
Opt. Lett. {\bf 40}, 4126-4129 (2015).

\bibitem{KMKLT} Y. V. Kartashov, B. A. Malomed, V. V. Konotop, V. E. Lobanov, and L. Torner, 
Opt. Lett. {\bf 40}, 1405-1408 (2015)
 
	\bibitem{Chiang95} K. S. Chiang,   
	Opt. Lett. {\bf 20}, 997 (1995).  

\bibitem{Wright}
E. M. Wright, G. I. Stegeman, and S. Wabnitz,
Phys. Rev. A {\bf 40}, 4455-4466 (1989). 
 
\bibitem{BenderBoet}  C. M. Bender and S. Boettcher, 
Phys. Rev. Lett. {\bf 80} 5243 (1998).


\bibitem{YK2}
H. Ramezani, T. Kottos, R. El-Ganainy, and D. N. Christodoulides, 
Phys. Rev. A \textbf{82}, 043803 (2010).

\bibitem{YK3}
A. A. Sukhorukov, Z. Xu, and Y. S. Kivshar, 
Phys. Rev. A \textbf{82}, 043818 (2010).

\bibitem{DribMal} R. Driben and B. A.  Malomed,  
Opt. Lett. {\bf 36}, 4323-4325 (2011).

\bibitem{AKOS} F. K. Abdullaev, V. V. Konotop, M. \"Ogren, and M. P. S\o rensen, 
Opt. Lett. {\bf 36}, 4566-4568 (2011).

\bibitem{review}  V. V. Konotop, J. Yang, and D. A. Zezyulin,
Rev. Mod. Phys.  {\bf 88}, 035002 (2016).

\bibitem{KKZ} Y. V. Kartashov,  V. V. Konotop, and D. A. Zezyulin,  
EPL (Europhysics Letters) 
{\bf 107}, 50002 (2014).

\bibitem{MalSak} H. Sakaguchi and B. A. Malomed,
New J. Phys. {\bf 18},   105005 (2016)


\bibitem{CP_optics} B. Dana, A. Bahabad, and B. A. Malomed,  
Phys. Rev. A {\bf 91}, 043808 (2015).


\bibitem{RWA} S. Mukamel, Principles of Nonlinear Optical Spectroscopy (Oxford University Press, 1995)

\bibitem{Barashenkov}  N. V. Alexeeva, I. V. Barashenkov, A. A. Sukhorukov, and Y. S. Kivshar, 
Phys. Rev. A {\bf 85}, 063837 (2012).
         
     
\end{thebibliography}
\end{document}